\documentclass[10pt]{iopart}

\usepackage{iopams}

\usepackage{graphicx}
\usepackage{hyperref}
\usepackage{amssymb}
\usepackage{longtable}
\usepackage{rotating}
\usepackage{color}
\usepackage{epsfig}
\usepackage{epsf}
\usepackage{fancyhdr}
\usepackage{enumerate}

\newcommand{\bea}{\begin{eqnarray}}
\newcommand{\eea}{\end{eqnarray}}

\newcommand{\be}{\begin{equation}}
\newcommand{\ee}{\end{equation}}

\newcommand{\F}{\mathcal{F}}

\newcommand{\To}{{T_\mathrm{obs}}}
\newcommand{\tav}[1]{\left\langle#1\right\rangle}

\newcommand{\mi}{\mathrm{i}}

\newcommand{\ve}{\varepsilon}

\begin{document}

\title[$\F$-statistic all-sky search for continuous gravitational waves in Virgo data]
{Implementation of an $\F$-statistic all-sky search for continuous gravitational waves in Virgo VSR1 data}
\author{LIGO Scientific Collaboration and Virgo Collaboration}

%% LSC authorlist in IOP format
%\documentclass[12pt]{iopart}
%\begin{document}
%\title{August 2013 LSC and September 2013 Virgo author list for the Fstat paper\\
%4-4-2014. IOP style}
\author{%
J.~Aasi$^{1}$,
B.~P.~Abbott$^{1}$,
R.~Abbott$^{1}$,
T.~Abbott$^{2}$,
M.~R.~Abernathy$^{1}$,
T.~Accadia$^{3}$,
F.~Acernese$^{4,5}$,
K.~Ackley$^{6}$,
C.~Adams$^{7}$,
T.~Adams$^{8}$
% ...
}

\address {$^{1}$LIGO - California Institute of Technology, Pasadena, CA 91125, USA }
\address {$^{2}$Louisiana State University, Baton Rouge, LA 70803, USA }
\address {$^{3}$Laboratoire d'Annecy-le-Vieux de Physique des Particules (LAPP), Universit\'e de Savoie, CNRS/IN2P3, F-74941 Annecy-le-Vieux, France }
\address {$^{4}$INFN, Sezione di Napoli, Complesso Universitario di Monte S.Angelo, I-80126 Napoli, Italy }
\address {$^{5}$Universit\`a di Salerno, Fisciano, I-84084 Salerno, Italy }
\address {$^{6}$University of Florida, Gainesville, FL 32611, USA }
\address {$^{7}$LIGO - Livingston Observatory, Livingston, LA 70754, USA }
\address {$^{8}$Cardiff University, Cardiff, CF24 3AA, United Kingdom }
% ...

%\end{document}

%\ead{krolak@impan.pl}

\begin{abstract}
We present an implementation of the $\F$-statistic to carry out the first search
in data from the Virgo laser interferometric gravitational wave detector
for periodic gravitational waves from {\em a priori} unknown, isolated rotating neutron stars.
We searched a frequency $f_0$ range from 100\,Hz to 1\,kHz
and the frequency dependent spindown $f_1$ range from $-1.6\,(f_0/100\,{\rm Hz}) \times 10^{-9}\,$\,Hz/s to zero.
A large part of this frequency - spindown space was unexplored by any of the all-sky searches published so far.
Our method consisted of a coherent search over two-day periods using the $\F$-statistic,
followed by a search for coincidences among the candidates from the two-day segments.
We have introduced a number of novel techniques and algorithms that allow the use of the Fast Fourier Transform
(FFT) algorithm in the coherent part of the search resulting in a fifty-fold speed-up in computation of
the $\F$-statistic with respect to the algorithm used in the other pipelines.
%that provide improvements in the efficiency compared to the other all-sky search pipelines.
No significant gravitational wave signal was found. The sensitivity of the search was estimated by injecting
signals into the data. In the most sensitive parts of the detector band more than 90\% of signals
would have been detected with dimensionless gravitational-wave amplitude greater than $5 \times 10^{-24}$.
\end{abstract}


%\maketitle

%%%%%%%%%%%%%%%%%%%%%%%%%%%%%%%%%%%%%%%%%%%%%%%%%%%%%%%%%%%%%%%%%%%%%%%%%%%%%%%%%%%%%%%%

\section{Introduction}
\label{sec:introduction}

This paper presents results from a wide parameter search for periodic gravitational waves from spinning neutron stars using data from the Virgo detector \cite{virgo:2012}.
The data used in this paper were produced during Virgo's first science run (VSR1)
which started on May 18, 2007 and ended on October 1, 2007.
The VSR1 data has never been searched for periodic gravitational waves
from isolated neutron stars before. The innovation of the search is the
combination of the efficiency of the FFT algorithm together with a nearly optimal grid of templates.

Rotating neutron stars are promising sources of gravitational radiation in the band
of ground-based laser interferometric detectors (see \cite{Abbott:2006vg} for a review).
In order for these objects to
emit gravitational waves, they must exhibit some asymmetry, which could be
due to imperfections of the neutron star crust or the influence of strong
internal magnetic fields (see \cite{MLM2013} for recent results). Gravitational radiation also may arise due to the r-modes,
i.e., rotation-dominated oscillations driven unstable by the gravitational emission (see \cite{O2010} for discussion of implications of r-modes for GW searches).
Neutron star precession is another gravitational wave emission mechanism (see \cite{JA2001} for a recent study).
Details of the above mechanisms of gravitational wave emission can be found
in \cite{NonAxNS1,NonAxNS2,NonAxNS3,NonAxNS4,NonAxNS5,JO2013}.

A signal from a rotating neutron star can be modeled independently of the specific mechanism
of gravitational wave emission as a nearly periodic wave the frequency of which decreases
slowly during the observation time due to the rotational energy loss. The signal registered by an Earth-based detector
is both amplitude and phase modulated because of the motion of the Earth with respect to the source.

The gravitational wave signal from such a star is expected to be very
weak, and therefore months-long segments of data must be analyzed.
The maximum deformation that a neutron star can
sustain, measured by the ellipticity parameter $\epsilon$ (Eq.\,(\ref{h0})),
ranges from $5 \times 10^{-6}$ for ordinary matter \cite{NonAxNS2,HK09} to $10^{-4}$
for strange quark matter \cite{NonAxNS5,JO2013}.
For unknown
neutron stars one needs to search a very large parameter space. As a result, fully coherent,
blind searches are computationally prohibitive. To perform a fully coherent search of VSR1 data
in real time (i.e., in time of 136 days of duration of the VSR1 run) over the parameter space proposed in this paper would require a $1.4 \times 10^4$ petaflop computer \cite{Brady1998a}.

A natural way to reduce the computational burden
is a hierarchical scheme, where first short segments of data are
analyzed coherently, and then results are combined incoherently.
This leads to computationally manageable searches at the expense of the signal-to-ratio loss.
To perform the hierarchical search presented in this paper in real time a $2.8$ teraflop computer is required.
One approach is to use short segments of the order of half an hour long so that the
the signal remains within a single Fourier frequency bin in each segment
and thus a single Fourier transform suffices to extract the whole power of the
gravitational wave signal in each segment.
Three schemes were developed for the analysis of Fourier transforms
of the short data segments and they were used in the all-sky searches
of ground interferometer data:  the ``stack-slide'' \cite{Abbott:2007tda},
the ``Hough transform'' \cite{HoughP1,Abbott:2007tda}, and
the ``PowerFlux'' methods \cite{Abbott:2007tda,Powerflux,AbadiePF2012}.

Another hierarchical scheme involves using longer segment duration
for which signal modulations need to be taken into account.
For segments of the order of days long, coherent analysis using
the popular $\F$-statistic \cite{JKSpaper} is computationally demanding,
but feasible. For example the hierarchical search of VSR1 data presented in this paper requires around
1900 processor cores for the time of 136 days which was the duration of the VSR1 run.
In the hierarchical scheme the first coherent analysis step is
followed by a post-processing step where data obtained in the first
step  are combined incoherently. This scheme was implemented by the
distributed volunteer computing project Einstein@Home~\cite{Einstweb}.
The E@H project performed analysis of LIGO S4 and S5 data, leading to results published
in three papers. In the first two, \cite{Abbott:2008uq,AbbottEHS5e2009},
the candidate signals from the coherent analysis were searched for coincidences.
In the third paper \cite{AasiEHS52013} the results from the coherent search were
analyzed using a Hough-transform scheme.

The search method used here is similar to the one used in the first two E@H
searches: it consists of coherent analysis by means of the $\F$-statistic, followed by a
search for coincidences among the candidates obtained in the coherent analysis. There are, however, important
differences: in this search we have a fixed threshold for the coherent analysis, resulting in a variable number of
candidates from analysis of each of the data segments. Moreover, for different bands we have a variable number
of two-day data segments (see Section \ref{sec:datasel} for details). In E@H searches the number of candidates
used in coincidences from each data segment was the same,
as was the number of data segments for each band. In addition, the duration of data segments for coherent
analysis in the two first E@H searches was 30h (48h in this case).

In this analysis we have implemented algorithms and techniques that
considerably improve the efficiency of this search. Most importantly
we have used the FFT
algorithm to evaluate the $\F$-statistic for two-day
data segments. Also we were able to use the FFT algorithm together with a grid of templates that was only
20\% denser than the best known grid (i.e., the one with the least number of points).

Given that the data we analyzed (Virgo VSR1) had a higher noise
and the duration was shorter than the LIGO S5 data
we have achieved a lower sensitivity than in the most recent E@H search \cite{AasiEHS52013}.
However due to very good efficiency of our code we were able to analyze a much larger parameter space.
We have analyzed a large part of the  $ f - \dot{f} $ plane that was previously unexplored
by any of the all-sky searches.
For example in this analysis the $ f - \dot{f} $ plane searched was $6.5$ larger than that
in the full S5  E@H search \cite{AasiEHS52013} (see Figure \ref{fig:ffdotpspace}).
In addition, the data from the Virgo detector is characterized by different spectral artifacts
(instrumental and environmental) from these seen in LIGO data. As a results, certain narrow bands excluded
from LIGO searches because of highly non-Gaussian noise can be explored in this analysis.

The paper is organized as follows. In Section \ref{sec:data} we describe the data from
the Virgo detector VSR1 run, in Section \ref{sec:datasel} we explain how the data were selected.
In Section \ref{sec:waveform} the response of the detector to a gravitational wave signal from a
rotating neutron star is briefly recalled. In Section \ref{sec:fstatistic} we introduce the $\F$-statistic.
In Section \ref{sec:pipeline} we describe the search method and present the algorithms for an efficient calculation
of the $\F$-statistic. In Section \ref{sec:veto} we describe the vetoing procedure of the candidates.
In Section \ref{sec:coincidences} we present the coincidence algorithm that
is used for post-processing of the candidates.
Section \ref{sec:search} contains results of the analysis. In Section \ref{sec:upperlimits}
we determine the sensitivity of this search, and we conclude the paper by Section \ref{sec:conclu}.
In \ref{a:coincidence} we present a general formula for probability
that is used in the estimation of significance of the coincidences.

\section{Data from the Virgo's first science run}
\label{sec:data}

\begin{figure}[h]
\begin{center}
\includegraphics[width=0.75\textwidth]{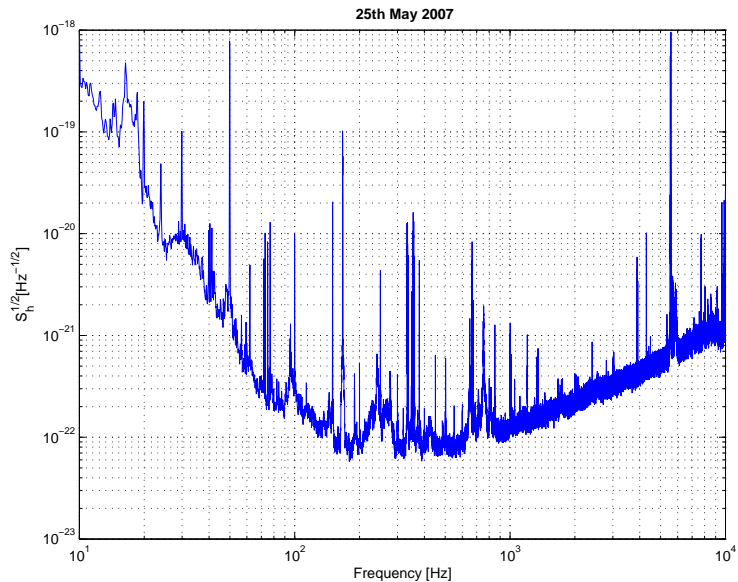}
\end{center}
\caption{Strain amplitude spectral density $\sqrt{S_h}$ of Virgo data
taken on May 25, 2007 during the VSR1 run.}
\label{fig:SensitivityH_VSR1_25May2007}
\end{figure}

The VSR1 science run spanned more than four months of data acquisition.
This run started on May 18, 2007 at 21:00 UTC and ended on October 1, 2007 at 05:00 UTC.
The detector was running close to its design sensitivity with a duty cycle of 81.0\%
\cite{VSR12008}. The data were calibrated, and the time series of the gravitational wave strain $h(t)$
was reconstructed. In the range of frequencies from 10Hz to 10kHz the systematic error in
amplitude was 6\% and the phase error was 70 mrad below the frequency of 1.9 kHz \cite{VSR1cal}.
A snapshot of the amplitude spectral density of VSR1 data in the Virgo detector band
is presented in Figure \ref{fig:SensitivityH_VSR1_25May2007}.
%
%%%%%%%%%%%%%%%%%%%%%%%%%%%%%%%%%%%%%%%%%%%%%%%%%%%%%%%%%%%%%%%%%%%%%%%%%%%%%%%%%%%%%%%%%%%%%%%%%%%%%%%%
\section{Data selection}
\label{sec:datasel}
\begin{figure}[ht]
\begin{center}
\includegraphics[width=0.75\textwidth]{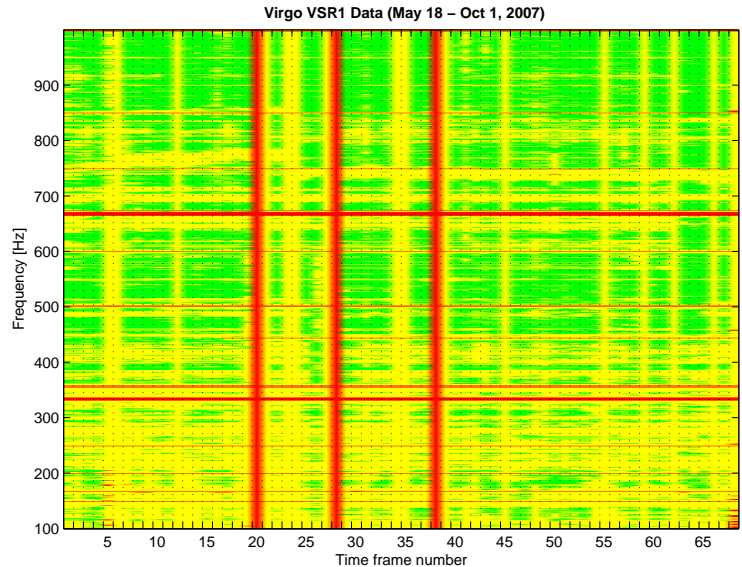}
\end{center}
\caption{A map of the VSR1 data. Grey - good data selected for the analysis,
white - data not analyzed because of large number of missing data or a strong variation
of the spectrum, black - bad data where number of missing data is greater than 50\%
or there are strong lines from mirror wires, electronics, etc.}
\label{fig:AllSkyVSR1}
\end{figure}

\begin{figure}[ht]
\begin{minipage}[b]{0.48\textwidth}
\centering
\includegraphics[angle=-90,width=\textwidth,clip]{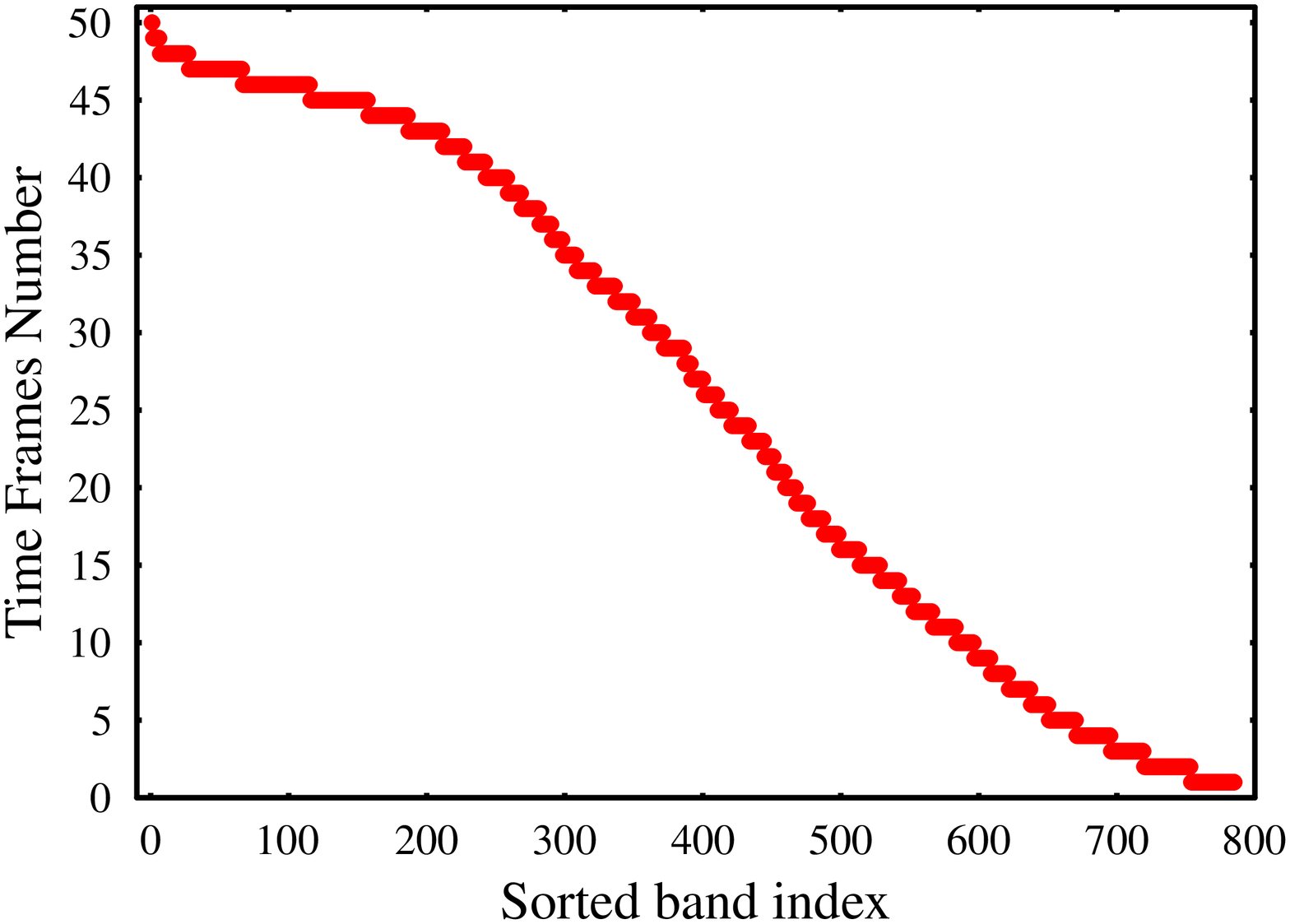}
%\caption{Number of good data segments. Upper panel - number of data segments
%for a given time frame. Lower panel - number of data segments for a given frequency band.}
%\label{fig:BandGood}
\end{minipage}
\hspace{0.04\textwidth}
\begin{minipage}[b]{0.48\textwidth}
\centering
\includegraphics[angle=-90,width=\textwidth,clip]{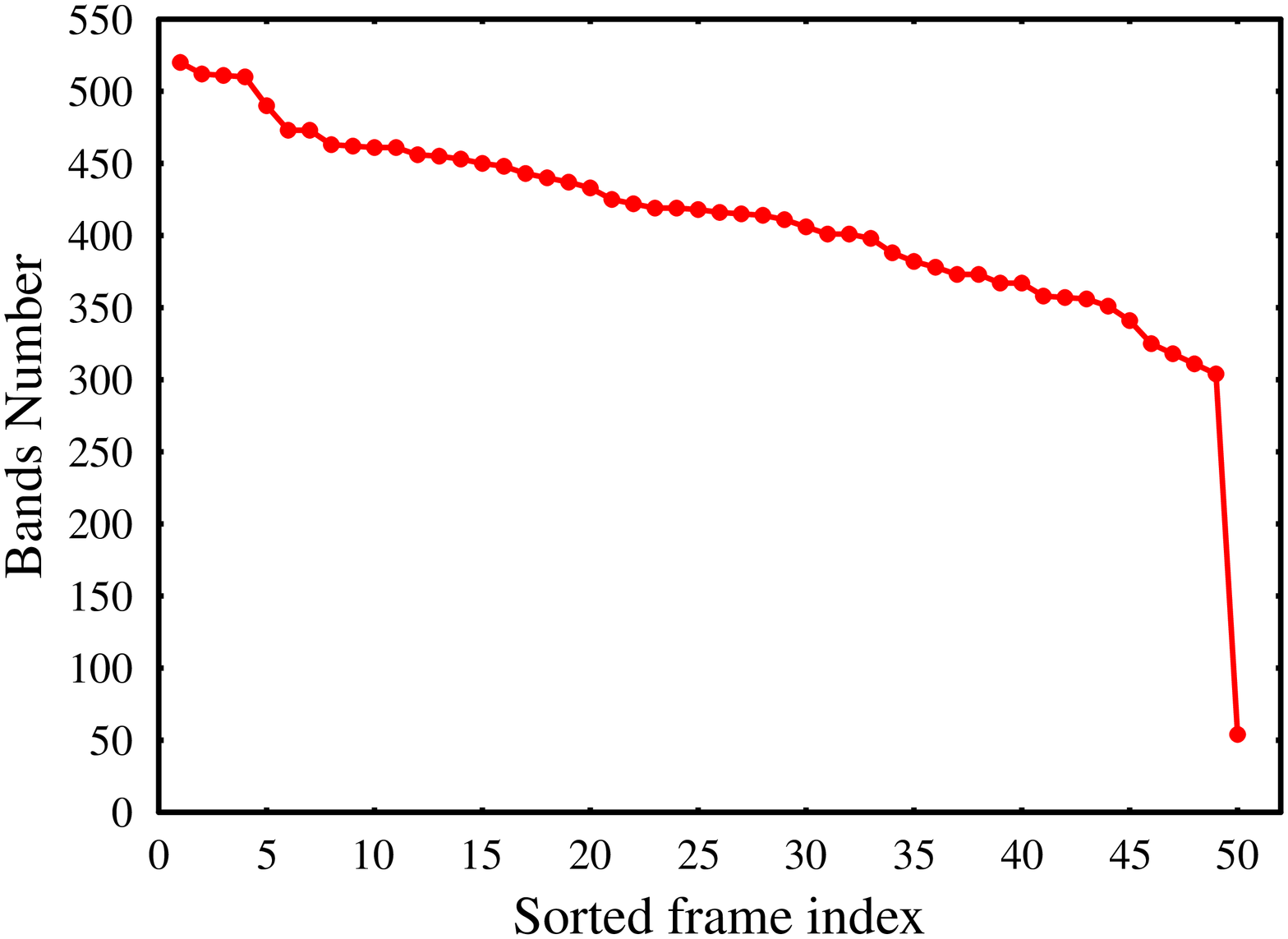}
%\label{fig:TimeGood}
\end{minipage}
\caption{Number of good data segments. Left panel - distribution of data segments
in frequency bands. The number of data segments chosen for the analysis in a given frequency band varies
from 1 to 50. Right panel - distribution of data segments time frames.
The number of data segments in a given time frame varies from 54 to 520.
The smallest, outlying number of bands of 54 is in the 5th time frame. In the remaining
frames the number of good bands is greater than 300.}
\label{fig:TimeBandGood}
\end{figure}

\begin{figure}[h]
\begin{center}
\includegraphics[width=0.75\textwidth]{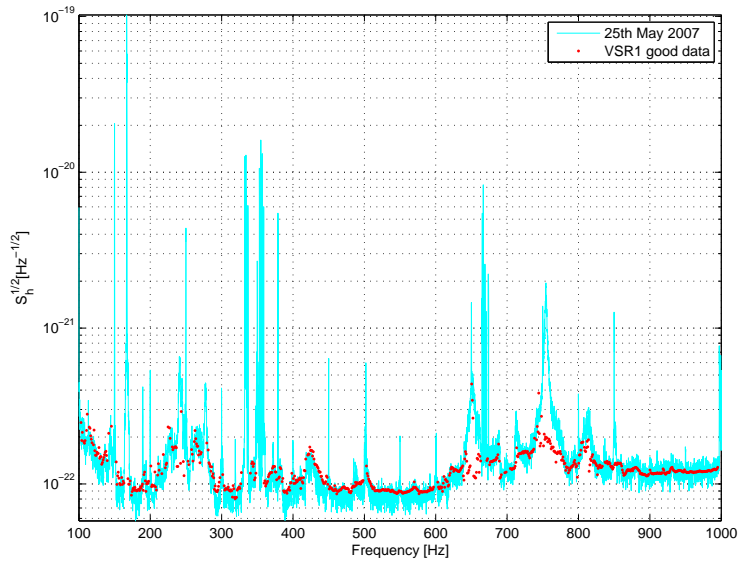}
\end{center}
\caption{A snapshot of strain amplitude spectral density $\sqrt{S_h}$ of the VSR1 data (grey curve)
in comparison with the spectral density estimated from the data used in the analysis (black dots).
The spectrum in each band is obtained
from a harmonic mean of the spectral densities of the data in each segment chosen for the analysis in the band.}
\label{fig:SensitivityH_VSR1_good}
\end{figure}

The analysis input consists of narrow-band
time-domain data segments. In order to obtain these sequences from VSR1 data, we have used the software
described in \cite{Astone2002}, and extracted the segments from the
Short Fourier Transform Database (SFDB).
The time domain data sequences were extracted
with a certain sampling time $\Delta t$, time duration $\To$ and offset frequency $f_{\rm off}$.
Thus each time domain sequence has the band $[f_{\rm off}, f_{\rm off} + B]$, where $B = 1/2 \times 1/\Delta t$.
We choose the segment duration
to be exactly two sidereal days and the sampling time equal to $1/2$\,s, i.e., the bandwidth $B = 1$~Hz.
As a result each narrow band time segment contains $N = 344\,656$ data points.
We have considered 67 two-day time frames for the analysis. The starting time of the analysis
(the time of the first sample of the data in the first time frame) was May 19, 2007 at 00:00 UTC.
The data in each time frame $d = 1,\dots,67$ were divided into narrow band sequences of 1Hz band each.
The bands overlap by $2^{-5}$~Hz resulting in 929 bands numbered from 0 to 928.
The relation between the band number $b$ and the offset frequency $f_{\rm off}$ is thus given by
\be
\label{eq:b2fo}
f_{\rm off} \mbox{[Hz]} = 100 + (1-2^{-5})\,b.
\ee
Consequently, we have 62\,243 narrow band data segments. From this set
we selected good data using the following conditions.
Let $N_0$ be the number of zeros in a given data segment (data point set to zero means that
at that time there was no science data).
Let $[f_{\rm off} , f_{\rm off} + B]$ be the band of a given data segment.
Let $S_{min}$ and  $S_{max}$ be the minimum and the maximum of the amplitude spectral
density in the interval $[f_{\rm off} + 0.05 B , f_{\rm off} + B -0.05 B]$.
Spectral density was estimated by dividing the data of a given segment into short stretches
and averaging spectra of all the short stretches.
We consider the data segment as good data and use it in this analysis, if the following two criteria are
met:
\begin{enumerate}[1.]
\item  $N_0/N < 1/4$,
\item  $S_{max}/S_{min} \leq 1.1$.
\end{enumerate}
20\,419 data segments met the above two criteria.
Figure \ref{fig:AllSkyVSR1} shows a time-frequency distribution of the good data
segments.
The good data appeared in 50 out of 67 two-day time frames and in 785 out of 929 bands.
The distribution of good data segments in time frames
and in bands is given in Figure \ref{fig:TimeBandGood}.
In Figure \ref{fig:SensitivityH_VSR1_good} we plot a snapshot spectral density of VSR1 data presented
in Figure \ref{fig:SensitivityH_VSR1_25May2007} and an estimate of the spectral density of the data
that was used in the analysis. We estimate the spectral density in each band by a harmonic mean
of the spectral densities of each of the two-day time segments chosen for the analysis in the band.
As the data spectrum in each band is approximately white, we estimate the spectral density in each time segment by
$2 \sigma^2 \Delta t$, where $\sigma^2$ is the variance of the data in a segment and $\Delta t$ is the
$1/2$s sampling time. We plot the band from 100Hz to 1 kHz which we have chosen for this search.
%

%%%%%%%%%%%%%%%%%%%%%%%%%%%%%%%%%%%%%%%%%%%%%%%%%%%%%%%%%%%%%%%%%%%%%%%%%%%%%%%%%%%%%%%%%%%%%%%%%%%%%%%%
\section{The response of the detector}
\label{sec:waveform}
The dimensionless noise-free response $h$ of a gravitational-wave detector
to a weak plane gravitational wave in the {\em long wavelength approximation}, i.e.,
when the characteristic size of the detector is much smaller than the reduced wavelength
$\lambda/(2\pi)$ of the wave,
can be written as the linear combination of the two independent wave polarizations $h_+$ and $h_\times$,
\be
\label{resp}
h(t) = F_+(t) h_+(t) + F_\times(t) h_\times(t),
\ee
where $F_+$ and $ F_\times$ are the detector's beam-pattern functions,
\bea
\label{patt}
F_+(t) = a(t)\cos 2\psi+b(t)\sin 2\psi,
\\[1ex]
F_\times(t) = b(t)\cos 2\psi-a(t)\sin 2\psi.
\eea
The beam-patterns $F_+$ and $ F_\times$ are linear combinations of $\sin2\psi$
and $\cos2\psi$, where $\psi$ is the polarization angle of the wave.
The functions $a(t)$ and $b(t)$ are the amplitude modulation functions,
that depend on the location and orientation of the detector on the Earth
and on the position of the gravitational-wave source in the sky,
described in the equatorial coordinate system
by the right ascension $\alpha$ and the declination $\delta$ angles.
They are periodic functions of time with the period of one and two sidereal days.
The analytic form of the functions $a(t)$ and $b(t)$ for the case of interferometric detectors
is given by Eqs.\ (12) and (13) of \cite{JKSpaper}. For a rotating nonaxisymmetric neutron star,
the wave polarization functions are of the form
\be
\label{hphc}
h_+(t) = h_{0+} \cos(\phi(t)+\phi_0),
\\[1ex]
h_\times(t) = h_{0\times} \sin(\phi(t)+\phi_0),
\ee
where $h_{0+}$ and $h_{0\times}$ are constant amplitudes of the two
polarizations and $\phi(t)+\phi_0$ is the phase of the wave,
$\phi_0$ being the initial phase of the waveform.
The amplitudes $h_{0+}$ and $h_{0\times}$ depend
on the physical mechanism responsible for the gravitational radiation,
e.g., if a neutron star is a triaxial ellipsoid rotating around
a principal axis with frequency $f$, then these amplitudes are
\be
\label{h0iota}
h_{0+} = \frac{1}{2} h_0 (1 + \cos^2\iota),
\\[1ex]
h_{0\times} = h_0 \cos\iota,
\ee
where $\iota$ is the angle between the star's angular momentum vector
and the direction from the star to the Earth, and the amplitude $h_0$ is given
by
\be
\label{h0}
h_0 = \frac{16\pi^2G}{c^4} \frac{\epsilon I f^2}{r}.
\ee
Here $I$ is the star's moment of inertia with respect to the rotation axis,
$r$ is the distance to the star, and $\epsilon$ is the star's ellipticity
defined by $\epsilon=|I_1-I_2|/I$, where $I_1$ and $I_2$ are moments of inertia
with respect to the principal axes orthogonal to the rotation axis.
We assume that the gravitational waveform given by Eqs.\ (\ref{resp})--(\ref{hphc})
is almost monochromatic around some angular frequency $\omega_0$,
which we define as instantaneous angular frequency
evaluated at the solar system barycenter (SSB) at $t=0$, and we assume
that the frequency evolution is accurately described by one spindown
parameter $\omega_1$. Then the phase $\phi(t)$ is given by
\begin{equation}
\label{eq:phase}
\phi(t) = \omega_0 t + \omega_1 t^2 + \frac{{\bf n} \cdot {\bf r}_d(t)}{c} (\omega_0 + 2 \omega_1 t),
\end{equation}
where, neglecting the relativistic effects, ${\bf r}_d(t)$ is the vector that joins
the SSB with the detector, and ${\bf n}$ is the unit vector pointing
from SSB to the source. In equatorial coordinates
$(\delta$, $\alpha)$ we have ${\bf n} = (\cos\delta\cos\alpha,\cos\delta\sin\alpha,\sin\delta)$.
%Approximations that lead to Eq.\ \ref{pha3} are discussed in detail
%in Sec.\ II~B and Appendix~A of Paper~I.

%%%%%%%%%%%%%%%%%%%%%%%%%%%%%%%%%%%%%%%%%%%%%%%%%%%%%%%%%%%%%%%%%%%%%%%%%%%%%%%%%%%%%%%%%
\section{The $\F$ - statistic}
\label{sec:fstatistic}
A method to search for gravitational wave signals from a rotating neutron star
in a detector data $x(t), t = 1, ... , N$
uses the $\F$-statistic, described in \cite{JKSpaper}.
The $\F$-statistic is obtained by maximizing the likelihood function with respect to the four unknown
parameters - $h_0$, $\phi_0$, $\iota$, and $\psi$. This leaves a function of
only the remaining four parameters - $\omega_0$, $\omega_1$, $\delta$, and $ \alpha$.
Thus the dimension of the parameter space that we need to search decreases from 8 to 4.
In this analysis we shall use an observation time $\To$ equal to the integer multiple of sidereal days.
Since the bandwidth of the signal over our coherent observation time of two days is very small,
we can assume that over this band the spectral density of the noise is white (constant).
Under these assumptions the  $\F$-statistic is given by \cite{JKpaper4,JKpaper5}
\be
\label{eq:Fstat}
\F \approx \frac{2}{\sigma^2}
\left( \frac{|F_a|^2}{\tav{a^2}} + \frac{|F_b|^2}{\tav{b^2}} \right),
\ee
where $\sigma^2$ is the variance of the data,
and
\bea
\label{Fab}
F_{a} := \sum^N_{t=1} x(t)\, a(t) \exp[-\mi\phi(t)],
\\ \nonumber
F_{b} := \sum^N_{t=1} x(t)\, b(t) \exp[-\mi\phi(t)].
\eea
\be
\tav{a^2} = \sum^N_{t=1} a(t)^2, \,\,\, \tav{b^2} = \sum^N_{t=1} b(t)^2.
\ee

%%%%%%%%%%%%%%%%%%%%%%%%%%%%%%%%%%%%%%%%%%%%%%%%%%%%%%%%%%%%%%%%%%%%%%%%%%%%%%%%%%%%%%%%%
\section{Description of the search}
\label{sec:pipeline}
The search consists of two parts; the first part is a coherent search of two-day data
segments, where we search a 4-parameter space defined by angular frequency
$\omega_0$, angular frequency derivative $\omega_1$, declination $\delta$, and right ascension
$\alpha$. The search is performed on a 4-dimensional grid in the parameter space
described in Section \ref{subsec:Fcal}. We set a fixed threshold of {\bf 20} for
the  $\F$-statistic for each data segment. This corresponds to a threshold of {\bf 6} for the signal-to-noise ratio.
All the threshold crossings are recorded together with corresponding 4 parameters of the grid point
and the signal-to-noise ratio $\rho$. The signal-to-noise is calculated
from the value of the $\F$-statistic at the threshold crossing as
\be
\label{eq:F2snr}
\rho = \sqrt{2 (\F - 2)}.
\ee
In this way for each narrow band segment we obtain a set of candidates. The candidates are then
subject to the vetoing procedure described in Section \ref{sec:veto}.
The second part of the search is the post-processing stage involving
search for coincidences among the candidates.
The coincidence procedure is described in Section \ref{sec:coincidences}.

\subsection{Choice of the parameter space}
\label{subsec:pspace}

We have searched the frequency band from 100~Hz to 1~kHz over the entire sky.
We have followed \cite{Brady1998a} to constrain the maximum value of the parameter
$\omega_1$ for a given frequency $\omega_0$ by $|\omega_1| \leq  \omega_0/(2 \tau_{min})$, where
$\tau_{min}$ is the minimum spindown age\footnote{The factor
of two in this formula appears here because the spindown parameter $\dot{f}$ used in
\cite{Brady1998a} is twice the spindown
parameter used in this work.}. We have
chosen $\tau_{min} = 1000$yr for the whole frequency
band searched. Also, in this search we have considered only the negative values for the parameter $\omega_1$,
thus assuming that the rotating neutron star is spinning down.
This gives the frequency-dependent range of the spindown parameter $f_1$
where $f_1 = \omega_1/(2\pi)$:
\be
\label{eq:sd}
|f_1| \leq  1.6\times 10^{-9} \frac{f}{100{\mbox Hz}} \frac{1000{\mbox yr}}{\tau_{min}}
[{\mbox Hz s}^{-1}]
\ee
We have considered only one frequency derivative. Estimates taking into account parameter correlations
(see \cite{Brady1998a} Figure 6 and Eqs. (6.2) - (6.6)) show that even for the minimum spindown age of $40$yr
and for two days coherent observation time that we consider here, it is sufficient to include
just one spindown parameter.

\begin{figure}
\begin{center}
\includegraphics[width=0.75\textwidth]{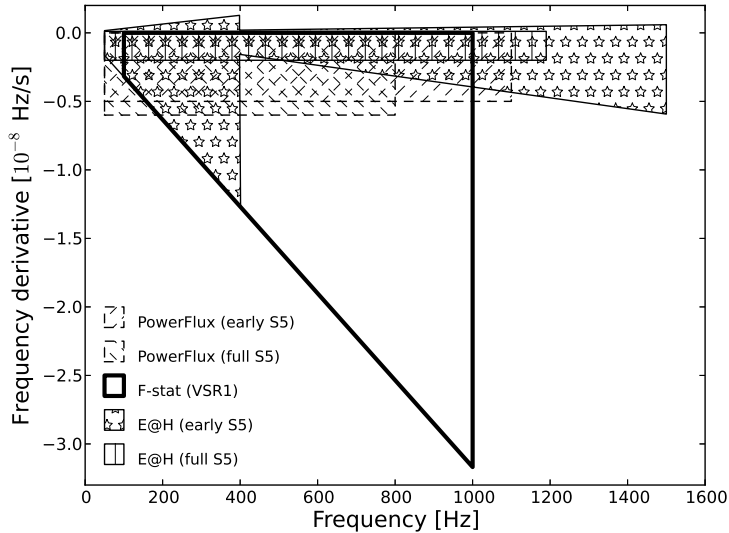}
\end{center}
\caption{Comparison of the parameter space in $ f - \dot{f} $ plane
searched in VSR1 analysis  presented in this paper (area enclosed by a thick black line)
and recently published PowerFlux and E@H searches of the LIGO S5 data.
\label{fig:ffdotpspace}}
\end{figure}

In Figure \ref{fig:ffdotpspace} we have compared the parameter space searched in this analysis
in the $f - \dot{f}$ plane with that of other recently published all-sky searches: Einstein@Home early S5 search
\cite{AbbottEHS5e2009}, Einstein@Home full S5 \cite{AasiEHS52013}, PowerFlux early S5 \cite{Collaboration:2008rg},
PowerFlux full S5 \cite{AbadiePF2012}.

\subsection{Efficient calculation of the $\F$-statistic on the grid in the parameter space}
\label{subsec:Fcal}
Calculation of the $\F$-statistic (Eq. (\ref{eq:Fstat})) involves two
sums given by Eqs. (\ref{Fab}).
By introducing a new time variable
called the barycentric time $t_b$ (\cite{schutz-91,JKSpaper,JKpaper5})
\be
\label{eq:BT}
t_b := t + \frac{{\bf n} \cdot {\bf r}_d(t)}{c}.
\ee
we can write these sums as discrete Fourier transforms in the following way
\bea
\label{Fabr}
F_{a} = \sum^N_{t_b=1} x(t(t_b))\, a(t(t_b)) \exp[-\mi\phi_s(t(t_b))] \exp[-\mi\omega_0 t_b ],
\\ \nonumber
F_{b} = \sum^N_{t_b=1} x(t(t_b))\, b(t(t_b)) \exp[-\mi\phi_s(t(t_b))] \exp[-\mi\omega_0 t_b ],
\eea
where
\be
\phi_s(t) = \omega_1 t^2 + 2 \frac{{\bf n} \cdot {\bf r}_d(t)}{c} \omega_1 t.
\ee
Written in this form, the two sums can be evaluated using the Fast Fourier Transform (FFT) algorithm
thus speeding up their computation dramatically. The time transformation described by equation
(\ref{eq:BT}) is called {\em resampling}.
In addition to the use of the FFT algorithm we apply an interpolation of the FFT using
the interbinning procedure (see \cite{JKpaper5} Section VB). This results in the $\F$-statistic
sampled twice as fine with respect to the standard FFT. This procedure is much faster than the interpolation
of the FFT obtained by padding the data with zeroes and calculating a FFT that is twice as long.
With the approximations described above for each value of the parameters $\omega_1$, $\delta$,
and $\alpha$, we calculate the $\F$-statistic efficiently for all the frequency bins in the
data segment of bandwidth 1Hz.

In order to search the 4-dimensional parameter space, we need to construct a 4-dimensional grid.
To minimize the computational cost we construct a grid that has the smallest number
of points for a certain assumed {\em minimal match} $MM$ \cite{MetricBen}. This problem is equivalent
to the {\em covering problem} \cite{conway-99,prix-07} and it has the optimal solution in 4-dimensions
in the case of a {\em lattice} i.e., a uniformly spaced grid. In order that our parameter space
is a lattice, the signals' reduced Fisher matrix must have components that are independent of the
values of the parameters. This is not the case for the signal given by Eqs.\,(\ref{resp}) - (\ref{eq:phase});
it can be realized however for an approximate signal called the {\em linear model}
described in Section IIIB of \cite{JKpaper5}.
The linear model consists of neglecting amplitude modulation of the signal and discarding
the component of the vector ${\bf r}_d(t)$ joining the detector
and the solar system barycenter that is perpendicular to the ecliptic.
This approximation is justified because the amplitude modulation is very slow compared to the phase
modulation and the discarded component in the phase is small compared to the others.
As a result the linear model signal $h_{lin}(t)$ has a constant amplitude $A_0$, and one can find
parameters such that the phases are linear functions of them.
We explicitly have (see Section IIIB of \cite{JKpaper5} for details):
\be
\label{eq:LinMod}
h_{lin}(t) = A_0\cos[\phi_{lin}(t) + \phi_0],
\ee
where
\be
\label{eq:LinPh}
\phi_{lin}(t) = \omega_0 t + \omega_1 t^2 + \alpha_1\mu_1(t) + \alpha_2\mu_2(t).
\ee
The parameters $\alpha_1$ and $\alpha_2$ are defined by
\bea
\label{eq:albe}
\alpha_1 := \omega_0 (\sin\alpha\cos\delta\cos\ve+ \sin\delta\sin\ve),
\\
\alpha_2 := \omega_0 \cos\alpha\cos\delta,
\eea
where $\ve$ is the obliquity of the ecliptic, and $\mu_1(t)$ and $\mu_2(t)$
are known functions of the detector ephemeris.

\begin{figure}[h]
\begin{center}
\includegraphics[width=0.75\textwidth]{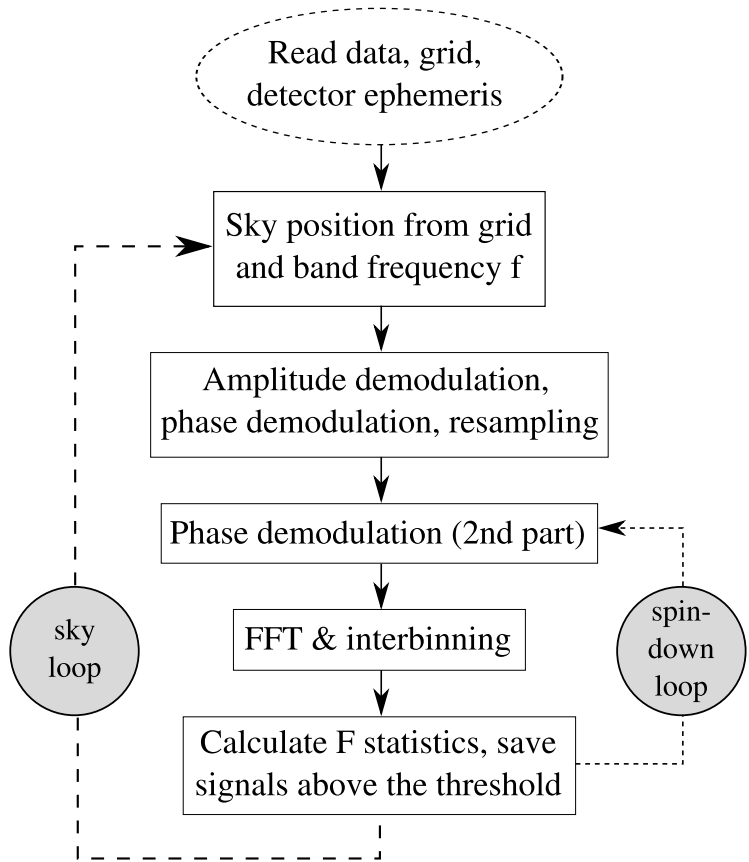}
\end{center}
\caption{Workflow of the $\F$-statistic search pipeline.}
\label{fig:flowdiagram}
\end{figure}

In order that the grid is compatible with application of the FFT, its points should be constrained
to coincide with Fourier frequencies at which the FFT is calculated. Moreover, we observed
that a numerically accurate
implementation of the interpolation to the barycentric time (see Eq.\,(\ref{eq:BT})) is so computationally
demanding that it may offset the advantage of the FFT. Therefore we introduced another constraint in
the grid such that the resampling is needed only once per sky position for all the spindown values.
Construction of the constrained grid is described in detail in Section IV of \cite{JKpaper5}.
In this search we have chosen the value of the minimal match $MM = \sqrt{3}/2$.
The workflow of the coherent part of the search procedure is presented in Figure \ref{fig:flowdiagram}.
\section{Vetoing procedure}
\label{sec:veto}
We apply three vetoing criteria to the candidates obtained in the coherent
part of the search - line width veto, stationary line veto and polar caps veto.
Data from the detector always contain some periodic interferences (lines) that are
detector artifacts. An important part of our vetoing procedure was to identify the lines in the data.
We have therefore performed a Fourier search with frequency resolution of
1/(2 days) $\simeq 5.8 \times 10^{-6}$~Hz for periodic signals of each of the two-day data
segments. We compared the frequencies of the significant periodic signals identified by our analysis with
the line frequencies obtained by the Virgo LineMonitor and we found that all the lines
from the LineMonitor were detected by our Fourier search.

\subsection{Line width veto}
We veto all the candidates with frequency $f$ around every known line frequency $f_l$
according to the following criterion
\begin{equation}
\label{eq:lveto}
|f - f_l| < \Delta f_{max},
\end{equation}
where the width $\Delta f_{max}$ is estimated as
\begin{equation}
\label{eq:lwidth}
\Delta f_{max} = f_l |v|_{max}/c + 2|f_1|_{max}(\To + \To |v|_{max}/c + |r|_{max}/c),
\end{equation}
where $v_{max}$ is the maximum value of the velocity of the detector
with respect to the SSB during the observation time, $r_{max}$ is the maximum distance to the SSB
during the coherent observation time $\To$, and $|f_1|_{max}$ is the maximum of the absolute value
of the frequency derivative.
Eq.\,(\ref{eq:lwidth}) determines the maximum smearing
of the frequency on each side of the line due to frequency modulation induced
by the filters applied in the $\F$-statistic.

\subsection{Stationary line veto}
Let us consider the instantaneous  frequency $f_{\rm inst}$ of the signal, i.e., the time derivative
of the phase:
\begin{equation}
\label{finst}
f_{\rm inst} := \frac{1}{2\pi}\frac{d\phi(t)}{dt} = f_0 + 2 f_1 t  +
            \frac{{\bf n} \cdot {\bf v}_d(t)}{c} (f_0 + 2 f_1 t) +
2\frac{{\bf n} \cdot {\bf r}_d(t)}{c} f_1,
\end{equation}
where ${\bf v}_d(t) = {d {\bf r}_d(t)}/{dt}$.
The frequency derivative of the instantaneous  frequency is given by
\begin{equation}
\label{finstdot}
\frac{d f_{\rm inst}(t)}{dt} = 2 f_1   + \frac{{\bf n} \cdot {\bf a}_d(t)}{c} (f_0 + 2 f_1 t) +
4\frac{{\bf n} \cdot {\bf v}_d(t)}{c} f_1,
\end{equation}
where ${\bf a}_d(t) = {d {\bf v}_d(t)}/{dt}$.
Eq.\,(\ref{finstdot}) is the rate of change of detector response frequency for a source whose
SSB frequency and spindown are $f_0$ and $f_1$. An instrumental line has a constant detector frequency
and mimics a source for which the r.h.s of Eq.\,(\ref{finstdot}) vanishes.
In practice, we veto candidates with
\begin{equation}
\left|2 f_1 + \frac{{\bf n} \cdot {\bf a}_d(t)}{c} f_0 +
2 \frac{{\bf n} \cdot {d^2 ({\bf r}_d(t)t)}/{dt^2}}{c} f_1\right| < \epsilon_{SL}
\end{equation}
for some $\epsilon_{SL}> 0$. In the search we choose $\epsilon_{SL} = 1/\To^2$, where
$\To$ is the observation time. The above  stationary line veto was introduced in reference \cite{Abbott:2007tda} and
refined in \cite{GlobCorr}; it was used in the first two E@H searches \cite{Abbott:2008uq,AbbottEHS5e2009}.

\subsection{Polar caps veto}
We observe that many of the detected lines cluster around the poles where declination $\delta$
is close to $\pm \pi/2$. An interference originating from a detector will correlate well with our
templates if the frequency modulation in Eq.\,(\ref{finst}) is minimized. Assuming that $f_1 = 0$
and that the diurnal motion of the Earth averages to 0 over two days observation time,
this happens when the quantity $|{\bf n} \cdot {\bf v}_d(t)|$ is minimized.
We find that this quantity is close to minimum independently of the value of $\alpha$ when $\delta = \pm \pi/2$.
Thus we veto candidates that are too close to the poles; we discard all candidates with
the declination angle $\delta$ within three grid cells from the poles.

\section{Coincidences}
\label{sec:coincidences}

In order to find coincidences among the candidates, we applied a method similar
to the one used in the first two E@H searches \cite{Abbott:2008uq,AbbottEHS5e2009}.
For each band we have searched for coincidences among candidates in different time frames.
We are able to search for coincidences only in those bands where there were two or more
time frames with data selected for the analysis.
If we search for a real gravitational wave signal, we must take into account
frequency evolution due to spindown of the rotating neutron star.
Thus the first step in the coincidence analysis
was to transform all frequencies $\omega_0(t_l)$ of the candidates
to a common fiducial reference time $t_f$. We have chosen the fiducial time to be
the time of the first sample of the latest time frame that we analyzed {\em i.e.}, the 67th
time frame. We have
\be
\label{eq:freqc}
\omega(t_f) = \omega_0(t_l) + 2\,\omega_1(t_l) [67\,\To  - t_l],
\ee
where $t_l$ is the time of the first sample of the $l$th time frame.
The next step was to divide the parameter space into cells.
This construction of the coincidence cell was
different from that in the E@H  analysis. To construct the cells in the parameter space
we have used the reduced Fisher matrix $\tilde{\Gamma}$ for the linear signal model
defined by Eqs.\,(\ref{eq:LinMod}) and (\ref{eq:LinPh}). The reduced Fisher matrix is the projected
Fisher matrix on the 4-dimensional space spanned by parameters
$\bf{\kappa} = (\omega_0, \omega_1, \alpha_1, \alpha_2)$.
We define the cell in the parameter space by the condition:
\be
\sum_{k,l}\tilde{\Gamma}_{kl}\kappa_k\kappa_l \leq 2.
\ee
Because the ephemeris of the detector is different in each of the time frames,
the reduced Fisher matrix is different in each time frame. To have a common
coincidence grid we have chosen the grid
defined by the latest frame i.e., the frame no. 67 as the coincidence grid.
After the transformation of the candidate frequencies to a reference time
and construction of the coincidence grid, the coincidence algorithm for each of the bands proceeded
in the following steps:
\begin{enumerate}[1.]
\item Transform angles $\alpha$ and $\delta$ to $\alpha_1$ and $\alpha_2$ coordinates (see Eq.(\ref{eq:albe}))

\item Transform candidate parameters to $x_l$ coordinates defined by
   \be
   x_l = \sum_{k=1}^4\kappa_k V_{kl}\sqrt{e_l},
   \ee
   where $V_{kl}, k = 1,..,4$ are eigenvectors of the matrix $\tilde{\Gamma}$,
   and $e_l$ are its eigenvalues. In these coordinates the Fisher matrix is
   proportional to the unit matrix.
\item Coordinates $x_l$ are rounded to the nearest integer.
   In this way we sort candidates efficiently into adjacent 4-dimensional hypercubes.
   If there are more than one candidates from a given data segment in a hypercube we select the
   candidate that has the highest SNR. We do sorting for each time frame in the band.
   If there is more than one candidate in a given hypercube we register a coincidence.
\item We shift cubes by 1/2 of their size in all possible $2^4$ directions, and for each
   shift we search for coincidences.
\end{enumerate}
This last step of the algorithm takes into account cases for which the candidate events are located
on opposite sides of cell borders, edges, and corners and consequently coincidences that could
not be found just by packing candidates into adjacent cells.

The most significant coincidence in each band is the one which has the highest multiplicity.
For each most significant coincidence we have calculated the false alarm probability {\em i.e.}, the
probability that such a coincidence can occur purely by chance. The false alarm probability
is calculated using the formula explained in \ref{a:coincidence}
and given by Eq.\,(\ref{eq:FAPs}).
This general formula applies to a variable number of candidates in various time slots and
also takes into account the $2^4$ shifts of the cells in the parameter space.

%%%%%%%%%%%%%%%%%%%%%%%%%%%%%%%%%%%%%%%%%%%%%%%%%%%%%%%%%%%%%%%%%%%%%%%%%%%%%%%%%%%%%%%%%
\section{The search}
\label{sec:search}

\begin{figure}[h]
\begin{center}
\includegraphics[width=30pc]{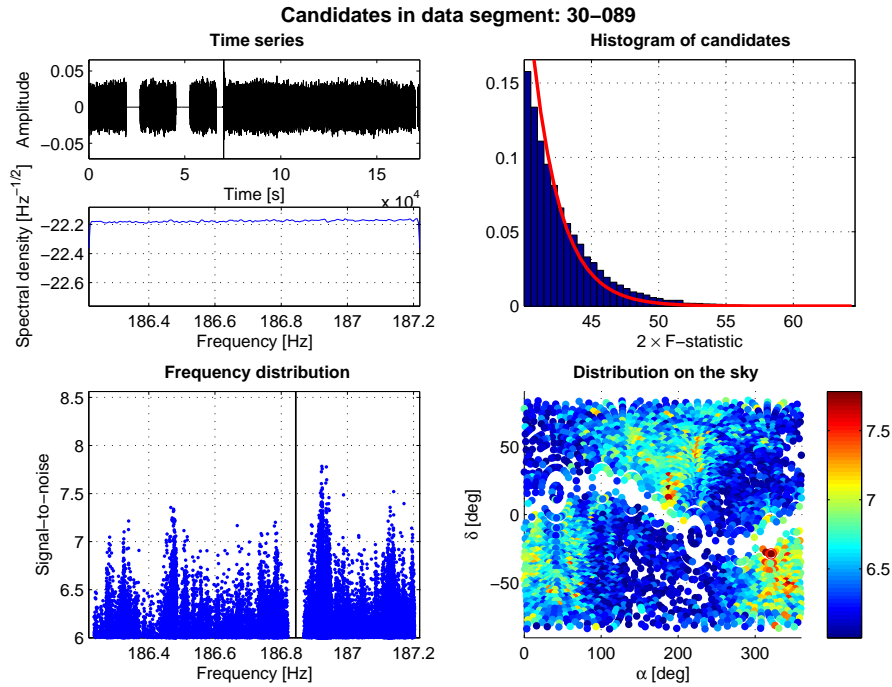}
\end{center}
\caption{Distribution of candidates after vetoing obtained in the coherent $\F$-statistic search of two-day data
segment $30-089$ i.e., a segment with the time frame number $d = 30$ and the band number $b = 89$.
The left top panel shows data and its spectrum. The right top panel shows the distribution of $2\F$ values of the candidates
in comparison to the $\chi^2$ distribution with 4 degrees of freedom. The bottom left panel shows
the distribution of signal-to-noise ratios $\rho$ (see Eq. (\ref{eq:F2snr}))
of the candidates as a function of the frequency. The
vertical black line is the periodic interference identified in the data. The candidates in the band around the
line are vetoed (see Eq.\,(\ref{eq:lwidth})). The right bottom panel shows the distribution of the candidates on the sky
in equatorial coordinates $\alpha$ and $\delta$. The distribution shows singularity at the ecliptic. This is a consequence
of the grid construction from the approximate linear signal model given by Eqs.(\ref{eq:LinMod}) - (\ref{eq:albe}).
\label{fig:candidates}}
\end{figure}

\begin{figure}[h]
  \begin{center}
  \includegraphics[width=0.75\textwidth]{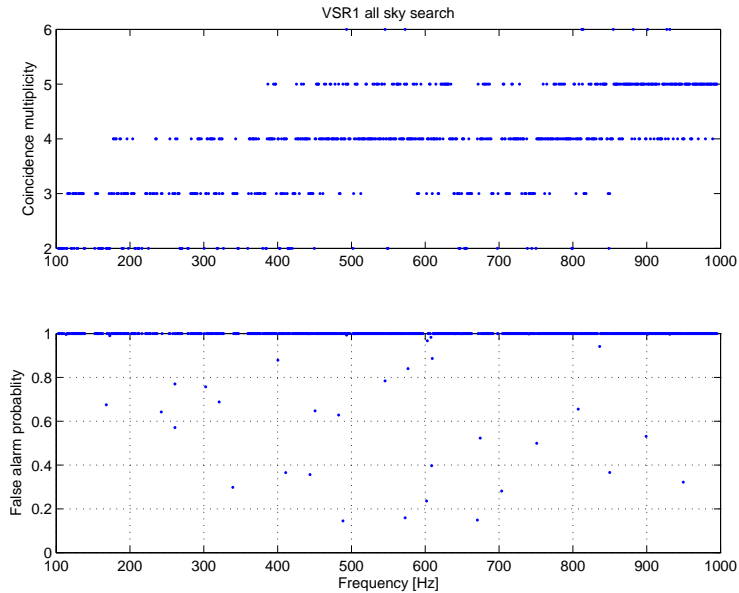}
  \caption{Coincidences among the candidates. Top panel: maximum multiplicity of coincidence
   as function of frequency. Bottom panel: corresponding coincidence false alarm probability.}
  \label{fig:Coin}
  \end{center}
\end{figure}

In this analysis we have searched coherently 20\,419 two-day time segments of data narrowbanded to
$1$~Hz. In the coherent part of the search described in Section \ref{sec:pipeline}
we have used $9.10\times 10^{16}$ templates which is the number of
$\F$-statistic values computed. This resulted in 20\,419 candidate files containing
$4.21\times 10^{10}$ candidates.
The candidates were subject to the vetoing using the three veto criteria:
line veto, polar caps veto, and stationary line veto described in Section \ref{sec:veto}.
As a result of vetoing around 24\% of the candidates were discarded leaving
$3.19\times 10^{10}$ candidates. Nearly all candidates were vetoed by the line
veto, whereas 0.20\% were vetoed by the stationary line criterion and only $3.2\times 10^{-2}$\%
by the polar caps veto.
In Figure \ref{fig:candidates} we present an example
of the candidate distribution obtained from the coherent search of one narrow band data segment
and after the vetoing procedure.
In the next step we have searched for the significant coincidences among the candidates.
We have searched for coincidences in all the frequency bands where there were two or more
data segments analyzed.  In Figure \ref{fig:Coin} we have plotted the highest coincidence
multiplicity for each of the bands. The highest multiplicity was 6, and it occurred in 10 bands.
The multiplicity tends to grow with the frequency, because the size of
the parameter space grows as $f^3$.
For each band we have calculated the false alarm
probability corresponding to the most significant coincidence using Eq.\,(\ref{eq:FAPs}).
The most significant coincidence occurred in band no.\ 401, corresponding
to the frequency range of $\sim [488.4, 489.4]$ Hz. It was a coincidence of multiplicity = 5,
and its false alarm probability was 14.5\%.
Note that a coincidence with the highest multiplicity is not the most significant.
This is because the significance depends on the number of time frames with candidates in a given band
and also on the number of candidates in the time frames.
By adopting a criterion used by E@H searches that the background
coincidences correspond to false alarm probability of 0.1\% or greater, we conclude that we have found
no significant coincidence and thus no viable gravitational wave candidate.
Considering the significance of the coincidences, we could adopt even a 10\%
false alarm probability as a background.
Consequently, we proceed to the final stage of our analysis - estimation of sensitivity
of the search.

%%%%%%%%%%%%%%%%%%%%%%%%%%%%%%%%%%%%%%%%%%%%%%%%%%%%%%%%%%%%%%%%%%%%%%%%%%%%%%%%%%%%%%%%%
\section{Sensitivity of the search}
\label{sec:upperlimits}

\begin{figure}[h]
  \begin{center}
  \includegraphics[width=0.75\textwidth]{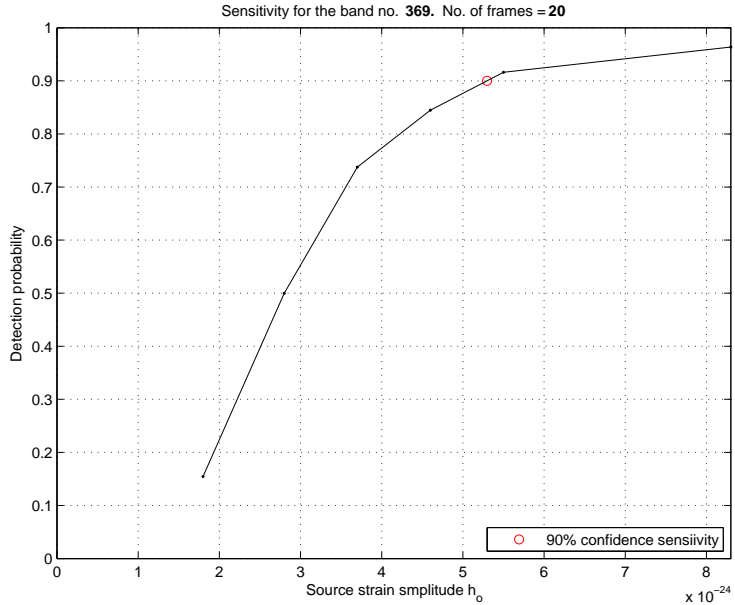}
    \end{center}
  \caption{Estimation of sensitivity in the band no. 369. In this band there are 20 time frames.
  Thus the signal is detected if it is coincident in more than 14 frames. The black dots are amplitudes
  of the injected signals and the corresponding detection probabilities estimated from the injections.
  The red circle is the interpolated amplitude corresponding to $90\%$ detection probability.
  This is the $90\%$ confidence sensitivity for this band.}
  \label{fig:senex}
\end{figure}

\begin{figure}[h]
  \begin{center}
  \includegraphics[width=0.75\textwidth]{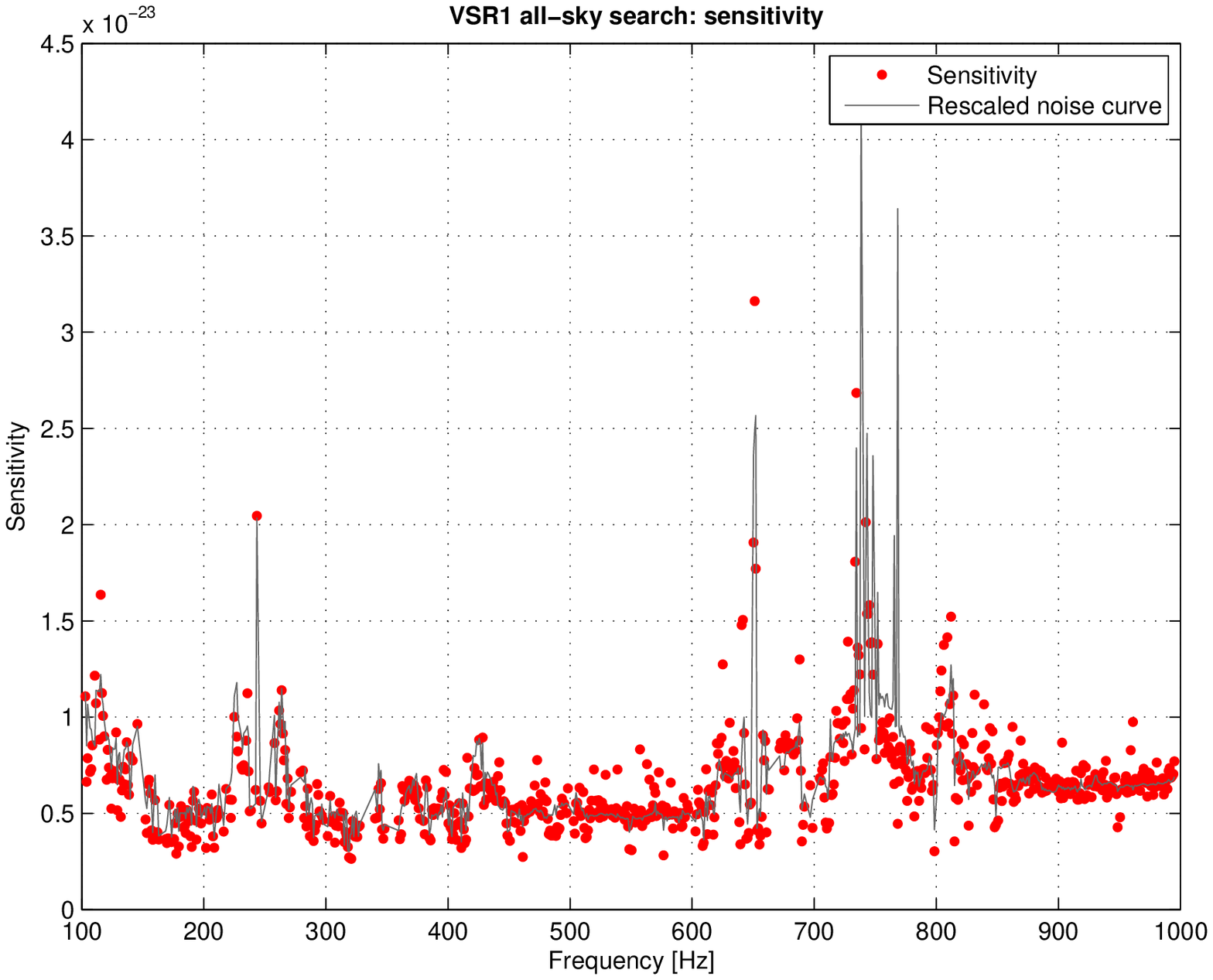}
  \end{center}
  \caption{The $90\%$ confidence sensitivity of the all-sky search of Virgo VSR1 data
           in the band from 100\,Hz to 1\,kHz. The dots show the source
           strain amplitude $h_0$ for which $90\%$ of sources are confidently detected
           by this pipeline. The thin line is the rescaled instrumental noise curve,
           see Eq.\,(\ref{eq:senspec}).}
  \label{fig:UL}
\end{figure}

The sensitivity of the search is taken to be the amplitude $h_0$ of the gravitational wave signal
that can be confidently detected. To estimate the sensitivity we use a procedure
developed in \cite{Abbott:2008uq}. We determine the sensitivity of the search in each of
the 785 frequency bands that we have searched. To determine the sensitivity,
we perform Monte Carlo simulations in which, for a given amplitude $h_0$, we randomly select the other
seven parameters of the signal: $\omega_0, \omega_1, \alpha, \delta, \phi_0, \iota,$ and $\psi$.
We choose frequency and spindown parameters uniformly over their range, and source positions uniformly over the sky.
We choose angles $\phi_0$ and $\psi$ uniformly over the interval $[0, 2\pi]$ and
we choose $\cos\iota$ uniformly over the interval $[-1, 1]$.
For each band we add the signal to all the data segments chosen for the analysis in that band.
Then we process the data through our pipeline. First, we perform a coherent $\F$-statistic search of each of the
data segments where the signal was added, and store all the candidates above our $\F$-statistic threshold  of 20.
In this coherent analysis, to make the computation manageable,  we search over only parameter space consisting
of $\pm$ 2 grid points around the nearest grid point where the signal was added. Then we apply our vetoing procedure
to the candidates obtained as explained in Section \ref{sec:veto}. Finally, we perform coincidence analysis
of the candidates that survive vetoing which is described in Section \ref{sec:coincidences}. We define a detectable signal
if it is coincident in more than 70\% of the
time frames in a given band. This condition is similar to the condition used in the two E@H
searches, where a coincidence method  was used \cite{Abbott:2008uq,AbbottEHS5e2009}.
For bands with only one frame available the coherent search over one $2$-day data segment was performed.
In this case the injected signal
is declared detected if its signal-to-noise ratio obtained in the coherent search is larger than the signal-to-noise
ratio of the loudest signal in that data segment without an injection.
For each band we inject signals with 5 different amplitude values, and perform
100 randomized injections for each amplitude.
For each amplitude we calculate how many signals were detected, and by interpolation we determine
the amplitude corresponding to 90\% of signals detected. This amplitude was defined as the 90\% confidence sensitivity.
Sometimes even for the highest amplitude we have not reached the $90\%$ detection probability. In this case  we performed
injections for higher amplitudes until the desired level of detectability was achieved.
In Figure \ref{fig:senex}, as an example, we present an estimation of the sensitivity in the band $b = 369$
corresponding to the frequency range of $\sim [457.47, 458.44]$ Hz.
The errors in the sensitivity estimates originate from calibration errors in the amplitude and errors due
to a finite number of Monte Carlo injections.
We use 100 injections; hence from a binomial statistic, one $\sigma$ is equivalent to 3\% fluctuation.
Thus the estimated amplitude sensitivity corresponds to confidence in the range from 87\% to 93\%.
To estimate how this uncertainty in confidence translates into uncertainty
in the amplitude we have performed an additional set of injections for a range of amplitudes
close to the estimated sensitivity and from the slope of the confidence vs. amplitude we determined
the uncertainty in the amplitude. To increase the accuracy of the error estimate, we have performed 1000
injections for each amplitude. The uncertainty in the amplitude was not more than 5\%.
The calibration errors in VSR1 data are 6\% (see Section \ref{sec:data}).
%and we estimated the Monte Carlo errors to be not more than 5\%.
Adding these two types of errors in quadrature results in the total error in sensitivity estimate to be around 7\%.
The sensitivity of this search obtained through Monte Carlo simulations for the whole band
searched is presented in Figure \ref{fig:UL}.
We see from Figure \ref{fig:UL} that the sensitivity essentially reflects the instrumental
noise curve given in Figure \ref{fig:SensitivityH_VSR1_good}. We have made a fit of the sensitivity
$h^{90\%}_0$ to the one-sided spectral density $S_h(f)$ of detector noise described
by the following relation:
\be
\label{eq:senspec}
h^{90\%}_0 = R_{\cal D}\sqrt{\frac{S_h(f)}{48~\mbox{hours}}}.
\ee
We find that the prefactor $R_{\cal D}$ is in the range from 15.6 to 22.4
and it depends on the frequency band and the number of the data segments in the band.

%Large parts of the VSR1 data band have a sensitivity in the range $5-6 \times 10^{-24}$. This
%range is better than in the case of the E@H LIGO S4 search \cite{Abbott:2008uq},
%where sensitivity was around $1 \times 10^{-23}$,
%it is comparable to the E@H early LIGO S5 search \cite{AbbottEHS5e2009} for the frequency band of $[800, 1000]$
%but it is worse in the most sensitive part of the E@H early LIGO S5 search band where sensitivity
%of $3 \times 10^{-24}$ was reached. Our sensitivity is markedly worse than the upper limits
%in the E@H full LIGO S5 data search \cite{AasiEHS52013} where in the band of $[100, 1000]$~Hz they were
%in the range of $0.8-3 \times 10^{-24}$.

%%%%%%%%%%%%%%%%%%%%%%%%%%%%%%%%%%%%%%%%%%%%%%%%%%%%%%%%%%%%%%%%%%%%%%%%%%%%%%%%%%%%%%%%%
\section{Conclusions}
\label{sec:conclu}

The sensitivity of this search was 50\% to 2 times better, depending on the
bandwidth, than that of the LIGO S4 search \cite{Abbott:2008uq}
and comparable to the sensitivities obtained in the early LIGO S5 data \cite{AbbottEHS5e2009},
but $2$ to $5$ times worse than the upper limits in the E@H full LIGO S5 data search \cite{AasiEHS52013}.
This was due to the lower noise and longer observation time for the LIGO S5 data w.r.t the Virgo VSR1 data.
However, for the first time in an all-sky search we have estimated the sensitivity
in the frequency band from 400Hz to 1kHz and the frequency spindown range from
$-7.2\,(f_0/400\,{\rm Hz}) \times 10^{-9}\,$\,Hz/s to $-6.0 \times 10^{-9}\,$\,Hz/s,
which is a previously unexplored region in the parameter space.

The next step is to test the search method described in this paper in the Mock Data Challenge (MDC) designed
by the LIGO and Virgo projects to validate and compare  pipelines that are proposed to be used in the analysis
of the forthcoming data  from the advanced detectors. It is also planned  to further test the pipeline presented
here with other data sets collected by the LIGO and Virgo detectors.

%As our search is very computationally efficient mainly due to the use of the FFT algorithm to evaluate the
%$\F$-statistic it constitutes a viable pipeline to be used in the quickly approaching advance detector era.

%\section{Acknowledgments}
\ack
%This research was supported in part by PL-Grid Infrastructure.
The authors gratefully acknowledge the support of the United States
National Science Foundation for the construction and operation of the
LIGO Laboratory, the Science and Technology Facilities Council of the
United Kingdom, the Max-Planck-Society, and the State of
Niedersachsen/Germany for support of the construction and operation of
the GEO600 detector, and the Italian Istituto Nazionale di Fisica
Nucleare and the French Centre National de la Recherche Scientifique
for the construction and operation of the Virgo detector. The authors
also gratefully acknowledge the support of the research by these
agencies and by the Australian Research Council, 
the International Science Linkages program of the Commonwealth of Australia,
the Council of Scientific and Industrial Research of India, 
the Istituto Nazionale di Fisica Nucleare of Italy, 
%%---- modified Feb2012:
% the Spanish Ministerio de Educaci\'on y Ciencia, 
the Spanish Ministerio de Econom\'ia y Competitividad,
%%----------------------
the Conselleria d'Economia Hisenda i Innovaci\'o of the
Govern de les Illes Balears, the Foundation for Fundamental Research
on Matter supported by the Netherlands Organisation for Scientific Research, 
the Polish Ministry of Science and Higher Education, the FOCUS
Programme of Foundation for Polish Science, the PL-Grid Infrastructure,
the Royal Society, the Scottish Funding Council, the
Scottish Universities Physics Alliance, The National Aeronautics and
Space Administration, the Carnegie Trust, the Leverhulme Trust, the
David and Lucile Packard Foundation, the Research Corporation, and
the Alfred P. Sloan Foundation.
This document has been assigned LIGO Laboratory document number LIGO-P1300133.

%%%%%%%%%%%%%%%%%%%%%%%%%%%%%%%%%%%%%%%%%%%%%%%%%%%%%%%%%%%%%%%%%%%%%%%%%%%%%%%%%%%%%%%%%
\section*{References}
\bibliography{FstatVSR1_CQG}

\appendix
\section{False alarm coincidence probability}
\label{a:coincidence}
Let us assume that for a given frequency band we analyze $L$ non-overlapping time segments.
Suppose that the search of the $l$th segment produces $N_l$ candidates.
Let us assume that the size of the parameter space for each time segment is the same,
and it can be divided into the number $N_{\rm cell}$ of independent cells.
We would like to test the null hypothesis that coincidences among
candidates from $L$ segments are accidental.
The probability for a candidate event to fall
into any given coincidence cell is equal to $1/N_{\rm cell}$.
Thus probability $\epsilon_l$ that a given coincidence cell
is populated with one or more candidate events is given by
\be
\epsilon_l = 1 - \Big(1 - \frac{1}{N_{\rm cell}}\Big)^{N_l}.
\ee
We may also consider independent candidates only, i.e., such that
there is no more than one candidate within one cell. If we obtain more
than one candidate within a given cell we choose the one which has
the highest signal-to-noise ratio. In this case
\be
\epsilon_l = \frac{N_l}{N_{\rm cell}}.
\ee
The probability $p_F(N_{\rm cell})$ that any given coincidence cell out of the total of
$N_{\rm cell}$ cells contains candidate events from
$C_{max}$ or more distinct data segments is given by a generalized
binomial distribution
\begin{eqnarray}
p_F(N_{\rm cell}) &=& \sum_{n=C_{max}}^{L} \frac{1}{n!(L-n)!} \times \nonumber \\
&\times& \sum_{\sigma\in\Pi(L)} \epsilon_{\sigma(1)}\ldots\epsilon_{\sigma(n)}
(1-\epsilon_{\sigma(n+1)})\ldots(1-\epsilon_{\sigma(L)}),
\end{eqnarray}
where $\sum_{\sigma \in \Pi(L)}$ is the sum over all the permutations of the $L$ data sequences.
Finally the probability $P_F$ that there is ${\mathcal C}_{max}$ or more coincidences
in one or more of the $N_{\rm cell}$ cells is
\be
\label{eq:FAcoin}
P_F = 1 - (1 - p_F(N_{\rm cell}))^{N_{\rm cell}}.
\ee
The above formula for the false alarm coincidence probability does not take into account
the case when candidate events are located on opposite sides of cell borders, edges, and corners.
In order to find these coincidences the entire cell coincidence grid is shifted
by half a cell width in all possible
$2^4 = 16$ combinations of the four parameter-space dimensions,
and coincidences are searched in all the
16 coincidence grids. This leads to a higher number of accidental
coincidences, and consequently Eq.\,\ref{eq:FAcoin} underestimates the false alarm
probability.  Let us consider the simplest one-dimensional case. In this case we have
$2^1 = 2$ possible shifts (the original coincidence grid and the one shifted by half).
This increases probability $p_F(N_{\rm cell})$ by a factor of $2$ if the two cell
coincidence grids were independent. However the cells overlap by half and some coincidences
would be counted twice. To account for this we divide the cells in the coincidence grid by
half resulting in $2 N_{\rm cell}$ cells and define the false alarm probability $p_F(2 N_{\rm cell})$
that any given half of the coincidence cell out of the total of
$2 N_{\rm cell}$ half cells contains candidate events from $C_{max}$ or more distinct data segments.
These are coincidences that were already counted, and consequently the false
alarm  probability with the cell shift is  $ 2 p_F(N_{\rm cell}) - p_F(2 N_{\rm cell}) $.
This results in the false alarm probability
\be
\label{eq:FAcoin1}
P_F = 1 - (1 - (  2 p_F(N_{\rm cell}) - p_F(2 N_{\rm cell}) ) )^{N_{\rm cell}}.
\ee
To generalize the above formula to higher dimensions we need to consider
further shifts and divisions of the cells. In the four dimension case this leads
to the formula for the probability $P^{\rm shifts}_F$ that there are ${\mathcal C}_{\mathrm{max}}$
or more independent coincidences in one or more of the $N_{\rm cell}$ cells in all
16 grid shifts given by
\bea
\label{eq:FAPs}
P^{\rm shifts}_F = 1 - \Big[ 1 - \Big( 2^4 p_F(N_c) \\ \nonumber
- \big( {4 \choose 1} p_F(2 N_c) + {4 \choose 2} p_F(2^2 N_c) +
    {4 \choose 3} p_F(2^3 N_c) + {4 \choose 4} p_F(2^4 N_c)   \big)   \\ \nonumber
- \big( {4 \choose 2} p_F(2^2 N_c) + {4 \choose 3} p_F(2^3 N_c) + {4 \choose 4} p_F(2^4 N_c) \big) \\ \nonumber
- \big( {4 \choose 3} p_F(2^3 N_c) + {4 \choose 4} p_F(2^4 N_c) \big) \\ \nonumber
-   {4 \choose 4} p_F(2^4 N_c)\Big)  \Big]^{N_c}
\eea

By choosing a certain false alarm probability $P_F$, we can calculate the
threshold number ${\mathcal C}_{\mathrm{max}}$ of coincidences. If we obtain more
than ${\mathcal C}_{\mathrm{max}}$ coincidences in our search we reject the null hypothesis
that coincidences are accidental only at the significance level of $P_F$.

\end{document}